\documentclass[Physsubmission, Phys]{SciPost}

\hypersetup{
    colorlinks,
    linkcolor={red!50!black},
    citecolor={blue!50!black},
    urlcolor={blue!80!black}
}

\usepackage[bitstream-charter]{mathdesign}
\DeclareSymbolFont{usualmathcal}{OMS}{cmsy}{m}{n}
\DeclareSymbolFontAlphabet{\mathcal}{usualmathcal}

\newcommand{\Pom}{\mathbb{P}}

\newcommand{\p}{\partial}
\newcommand{\twosidep}[1]{\stackrel{\leftrightarrow}{\p}_{\! #1}}

\newcommand*\wideestimates{\mathrel{\widehat{=}}}

\begin{document}

\begin{center}{\Large \textbf{
Central exclusive diffractive production of axial-vector $f_{1}$ mesons in proton-proton collisions\\
}}\end{center}

\begin{center}
Piotr Lebiedowicz\textsuperscript{1 $\star$},
Josef Leutgeb\textsuperscript{2},
Otto Nachtmann\textsuperscript{3},
Anton Rebhan\textsuperscript{2} and
Antoni Szczurek\textsuperscript{1}
\end{center}

\begin{center}
{\bf 1} 
Institute of Nuclear Physics Polish Academy of Sciences, Radzikowskiego 152, PL-31342 Krak{\'o}w, Poland
\\
{\bf 2} Institut f\"ur Theoretische Physik, Technische Universit\"at Wien,
Wiedner Hauptstrasse 8-10, A-1040 Vienna, Austria
\\
{\bf 3} Institut f\"ur Theoretische Physik, Universit\"at Heidelberg,
Philosophenweg 16, D-69120 Heidelberg, Germany
\\
* Piotr.Lebiedowicz@ifj.edu.pl
\end{center}



\definecolor{palegray}{gray}{0.95}
\begin{center}
\colorbox{palegray}{
  \begin{tabular}{rr}
  \begin{minipage}{0.1\textwidth}
    \includegraphics[width=22mm]{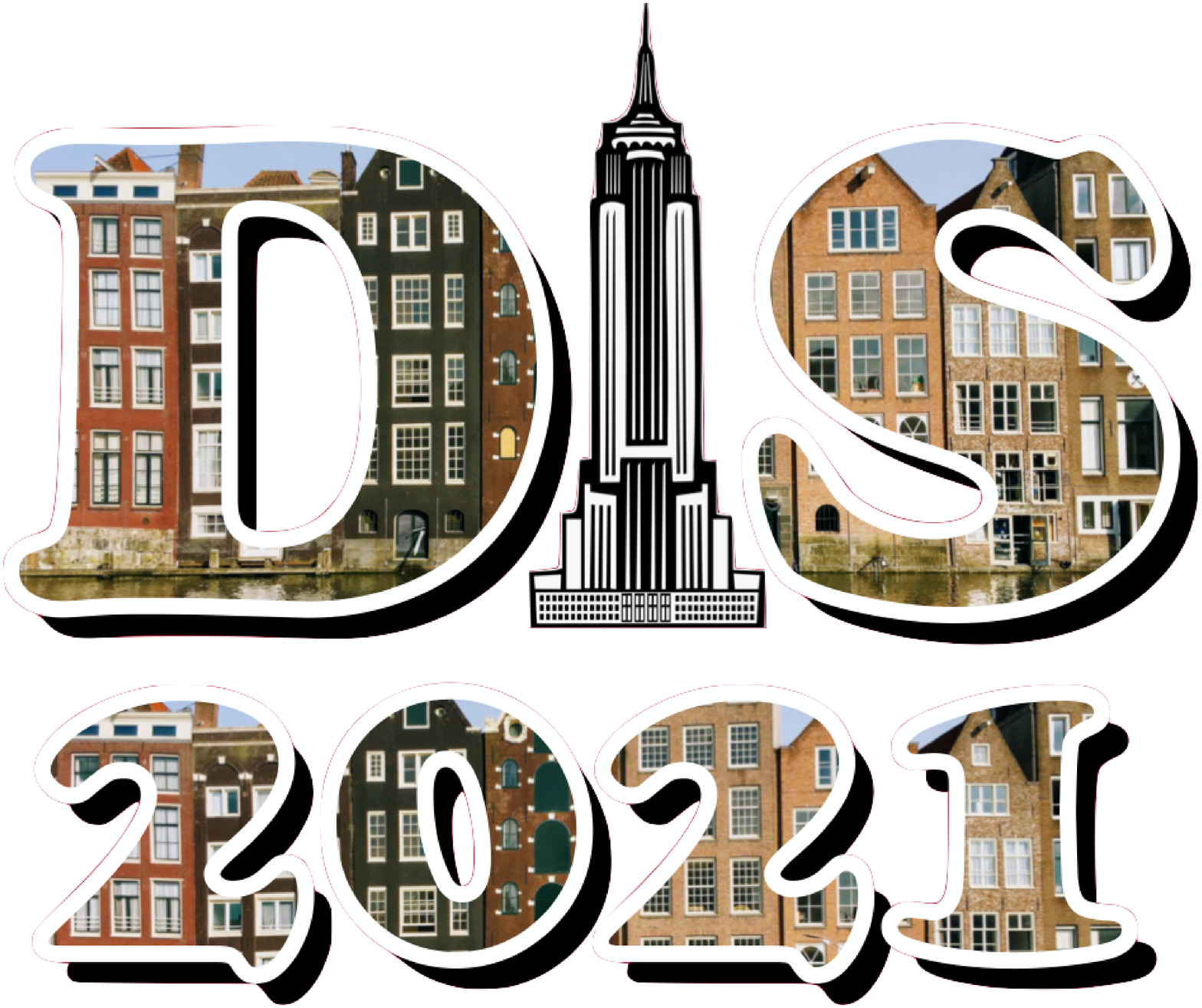}
  \end{minipage}
  &
  \begin{minipage}{0.75\textwidth}
    \begin{center}
    {\it Proceedings for the XXVIII International Workshop\\ on Deep-Inelastic Scattering and
Related Subjects,}\\
    {\it Stony Brook University, New York, USA, 12-16 April 2021} \\
    \doi{10.21468/SciPostPhysProc.?}\\
    \end{center}
  \end{minipage}
\end{tabular}
}
\end{center}

\section*{Abstract}
{\bf
Exclusive production of axial-vector $f_{1}(1285)$ meson
in proton-proton collisions via pomeron-pomeron fusion 
within the tensor-pomeron approach is discussed.
Two ways to construct the pomeron-pomeron-$f_{1}$ coupling
are presented.
We adjust the parameters of our model to the WA102 experimental data and compare with predictions of the Sakai-Sugimoto model.
Predictions for LHC experiments are given.
}


\section{Introduction}
\label{sec:intro}
\begin{figure}[h]
\centering
\includegraphics[width=.3\textwidth]{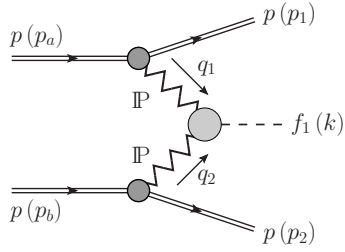}
\caption{Diagram for the reaction (\ref{1.1}) 
with double-pomeron exchange.}
\label{fig1}
\end{figure}

In this contribution we discuss \underline{central exclusive
production (CEP)} of $f_{1}$ ($J^{PC} = 1^{++}$) mesons
in proton-proton collisions
\begin{equation}
p\,(p_{a}) + p\,(p_{b}) \to p\,(p_{1}) + f_{1}\,(k) + p\,(p_{2}).
\label{1.1}
\end{equation}
As a concrete example we shall consider the $f_{1}(1285)$ meson.
The presentation is based on \cite{Lebiedowicz:2020yre} where
all details and many more results can be found.
At high energies the $\Pom \Pom$ fusion process (figure~\ref{fig1})
is expected to be dominant.
The relevant kinematic quantities are
\begin{eqnarray}
s = (p_{a} + p_{b})^{2},\;
q_1 = p_{a} - p_{1}, \;q_2 = p_{b} - p_{2},\; k = q_{1} + q_{2},\;
t_1 = q_{1}^{2}, \;t_2 = q_{2}^{2},\; m_{f_{1}}^{2} = k^{2}.
\label{1.2}
\end{eqnarray}
%

We treat our reaction in the \underline{tensor-pomeron approach}
as introduced in \cite{Ewerz:2013kda}.
This approach has a good basis from nonperturbative QCD
using functional integral techniques \cite{Nachtmann:1991ua}.
We describe the charge-conjugation $C = +1$ pomeron 
as effective rank 2 symmetric tensor exchange.
A tensor character of the pomeron is also preferred
in holographic QCD.

There are by now many applications of the tensor-pomeron model
to two-body hadronic reactions \cite{Ewerz:2016onn},
to photoproduction,
to DIS structure functions at low $x$, and especially to CEP reactions
$p + p \to p + X + p$, where
$X = \eta,\, \eta',\, f_{0},\, f_{2},\, \pi^{+}\pi^{-},\,
K\bar{K},\, p \bar{p},\, 4\pi,\, 4K,\, \rho^{0},\, \phi,\, \phi \phi,\, K^{*0} \bar{K}^{*0}$;
see e.g. \cite{Lebiedowicz:2013ika,Lebiedowicz:2016ioh,Lebiedowicz:2019por,Lebiedowicz:2019boz,Lebiedowicz:2019jru,Lebiedowicz:2021pzd}.

\section{Brief overview of the formalism}
\label{sec:formalism}

\subsection{The amplitude for the $pp \to pp f_{1}$ reaction}

The Born-level $\Pom\Pom$-fusion amplitude for the reaction (\ref{1.1})
can be written as
\begin{eqnarray}
{\cal M}^{(\Pom \Pom \to f_{1})}_{\lambda_{a} \lambda_{b} \to \lambda_{1} \lambda_{2} \lambda}
&=& (-i)\, (\epsilon^{\mu}(\lambda))^{*}\,
\bar{u}(p_{1}, \lambda_{1}) 
i\Gamma^{(\Pom pp)}_{\mu_{1} \nu_{1}}(p_{1},p_{a}) 
u(p_{a}, \lambda_{a}) \nonumber \\
&& \times 
i\Delta^{(\Pom)\, \mu_{1} \nu_{1}, \alpha_{1} \beta_{1}}(s_{1},t_{1}) \,
i\Gamma^{(\Pom \Pom f_{1})}_{\alpha_{1} \beta_{1}, \alpha_{2} \beta_{2}, \mu}(q_{1},q_{2}) \,
i\Delta^{(\Pom)\, \alpha_{2} \beta_{2}, \mu_{2} \nu_{2}}(s_{2},t_{2}) \nonumber \\
&& \times 
\bar{u}(p_{2}, \lambda_{2}) 
i\Gamma^{(\Pom pp)}_{\mu_{2} \nu_{2}}(p_{2},p_{b}) 
u(p_{b}, \lambda_{b}) \,.
\label{amplitude_f1_pompom}
\end{eqnarray}
Here $\epsilon^{\mu}(\lambda)$ is the polarisation vector of the $f_{1}$ meson,
$\Delta^{(\Pom)}$ and $\Gamma^{(\Pom pp)}$ 
denote the effective propagator and proton vertex function, respectively, 
for the tensor-pomeron exchange \cite{Ewerz:2013kda}.
The new quantity, to be studied here, is the $\Pom \Pom f_{1}$ coupling.
In practice we work with the amplitudes in the high-energy approximation.
In our analysis we include absorptive corrections
within the one-channel-eikonal approach.

\subsection{The pomeron-pomeron-$f_{1}$ coupling}




We follow two strategies for constructing 
the $\Pom \Pom f_{1}$ coupling and the vertex
function.

\textbf{(1)} Phenomenological approach.
First we consider a fictitious process: the fusion of two
``real spin-2 pomerons'' (or tensor glueballs) of mass $m$
giving an $f_{1}$ meson of $J^{PC} = 1^{++}$.
We make an angular momentum analysis of this reaction 
in its c.m. system, the rest system of the $f_{1}$ meson:
$\Pom\,(m, \epsilon_{1}) + \Pom\,(m, \epsilon_{2}) \to
f_{1}\,(m_{f_{1}},\epsilon)$.
The spin~2 of these ``pomerons'' can be combined to a total spin $S$
($0 \leqslant S \leqslant 4$) and this must be combined with
the orbital angular momentum $l$ to give the $J^{PC} = 1^{++}$
values of the $f_{1}$.
There are two possibilities,
$(l,S) = (2,2)\; \rm{and}\; (4,4)$
(see Appendix~A of \cite{Lebiedowicz:2013ika}),
and corresponding bare coupling Lagrangians $\Pom \Pom f_{1}$ are:
\begin{eqnarray}
&&
{\cal L}^{(2,2)}_{\Pom \Pom f_{1}} = \frac{g'_{\Pom \Pom f_{1}}}{32\,M_{0}^{2}}
\Big( \Pom_{\kappa \lambda} 
\twosidep{\mu} \twosidep{\nu}
\Pom_{\rho \sigma} \Big)
\Big( \p_{\alpha} U_{\beta} - \p_{\beta} U_{\alpha} \Big)\,
\Gamma^{(8)\,\kappa \lambda, \rho \sigma, \mu \nu, \alpha \beta}\,,
\label{2.3}\\
&&
{\cal L}^{(4,4)}_{\Pom \Pom f_{1}} = \frac{g''_{\Pom \Pom f_{1}}}{24 \cdot 32 \cdot M_{0}^{4}}
\Big( \Pom_{\kappa \lambda}
\twosidep{\mu_{1}} \twosidep{\mu_{2}} \twosidep{\mu_{3}} \twosidep{\mu_{4}}
\Pom_{\rho \sigma} \Big)
\Big( \p_{\alpha} U_{\beta} - \p_{\beta} U_{\alpha} \Big)\,
\Gamma^{(10)\,\kappa \lambda, \rho \sigma, \mu_{1} \mu_{2} \mu_{3} \mu_{4}, \alpha \beta}\,, \;\;\quad
\label{2.4}
\end{eqnarray}
where $M_{0} \equiv 1$~GeV (introduced for dimensional reasons),
$g'_{\Pom \Pom f_{1}}$ and $g''_{\Pom \Pom f_{1}}$ are
dimensionless coupling constants, 
$\Pom_{\kappa \lambda}$ is the $\Pom$ effective field,
$U_{\alpha}$ is the $f_{1}$ field,
and $\Gamma^{(8)}$, $\Gamma^{(10)}$ 
are known tensor functions~\cite{Lebiedowicz:2020yre}.
We use then these couplings, supplemented by suitable
form factors, for the $f_{1}(1285)$ CEP reaction (\ref{1.1}).

\textbf{(2)} Our second approach uses holographic QCD, in particular
the Sakai-Sugimoto model \cite{Sakai:2004cn,Brunner:2015oqa}
where the $\Pom \Pom f_{1}$ coupling is determined 
by the mixed axial-gravitational anomaly of QCD.
In this approach
\begin{eqnarray}
{\cal L}^{\rm CS} = \varkappa' \,U_{\alpha}\,\varepsilon^{\alpha \beta \gamma \delta}\,
\Pom^{\mu}_{\;\;\beta}\, \p_{\delta}\Pom_{\gamma \mu}
+ \varkappa'' \,U_{\alpha}\,\varepsilon^{\alpha \beta \gamma \delta}\,
\left( \p_{\nu}\Pom^{\mu}_{\;\;\beta} \right) 
\left( \p_{\delta}\p_{\mu}\Pom^{\nu}_{\;\;\gamma} - \p_{\delta}\p^{\nu}\Pom_{\gamma \mu} \right)
\label{2.5}
\end{eqnarray}
with $\varkappa'$ a dimensionless constant and
$\varkappa''$ a constant of dimension GeV$^{-2}$; see Appendix~B of \cite{Lebiedowicz:2020yre}.

For our fictitious reaction ($\Pom + \Pom \to f_{1}$) there is
strict equivalence
${\cal L}^{{\rm CS}} \wideestimates 
{\cal L}^{(2,2)} + {\cal L}^{(4,4)}$
if the couplings satisfy the relations
\begin{eqnarray}
g'_{\Pom \Pom f_{1}} =
-\varkappa'\,\frac{M_{0}^{2}}{k^{2}}
-\varkappa''\,\frac{M_{0}^{2}(k^{2}-2 m^{2})}{2k^{2}} \,,
\qquad
g''_{\Pom \Pom f_{1}} =
\varkappa''\,\frac{2 M_{0}^{4}}{k^{2}} \,.
\label{2.7}
\end{eqnarray}
For our CEP reaction (\ref{1.1}) we are dealing with pomerons of mass
squared $t_{1}, t_{2} < 0$ and, in general, $t_{1} \neq t_{2}$.
Then, the equivalence relation
for small values $|t_{1}|$ and $|t_{2}|$
will still be approximately true and we confirm this
by explicit numerical studies 
(see Fig.~11 of \cite{Lebiedowicz:2020yre}).

\section{Results}
\label{sec:results}

\subsection{Comparison with the WA102 data}

\begin{figure}[!ht]
\centering
\includegraphics[width=0.4\textwidth]{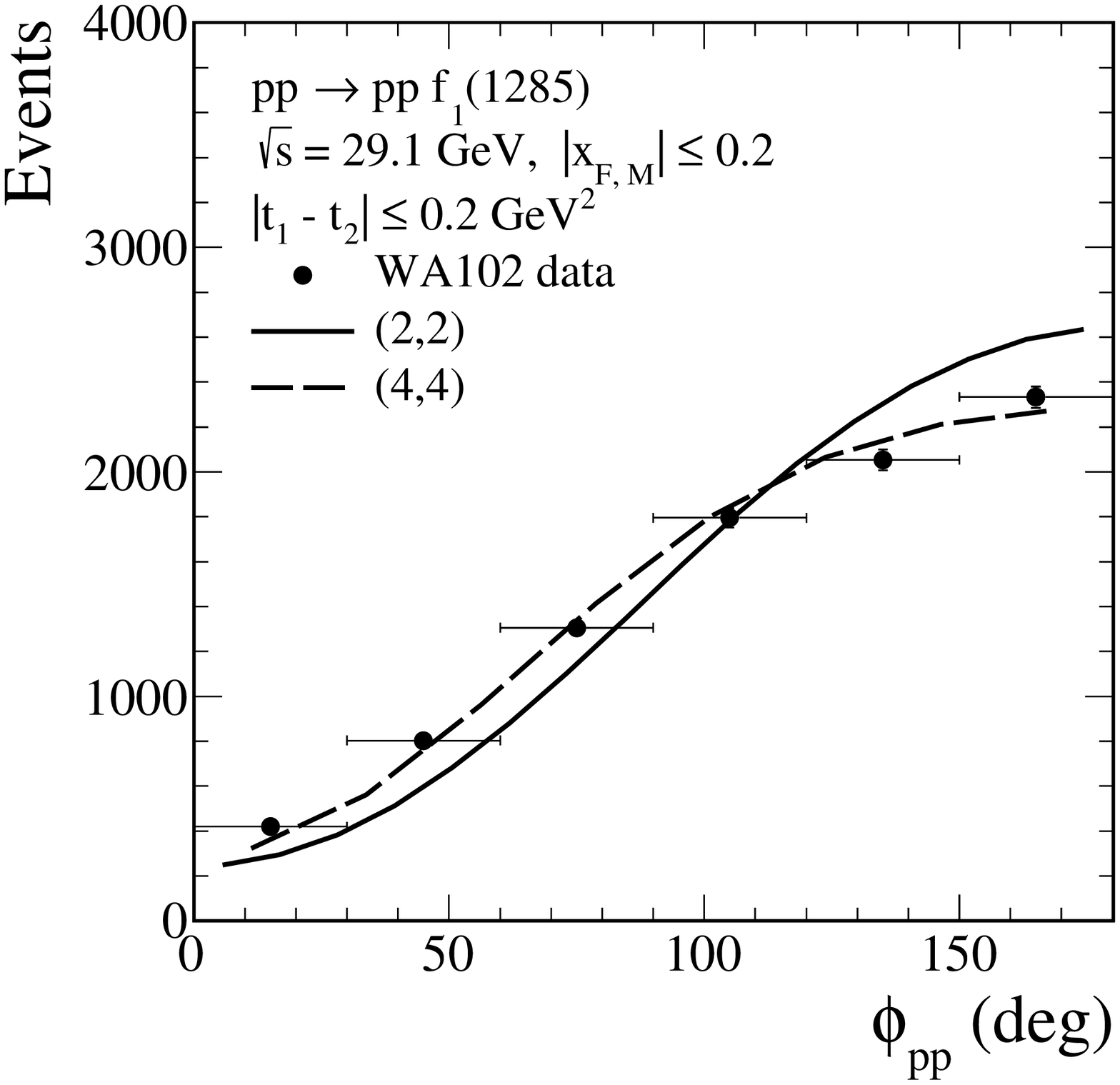}
\includegraphics[width=0.4\textwidth]{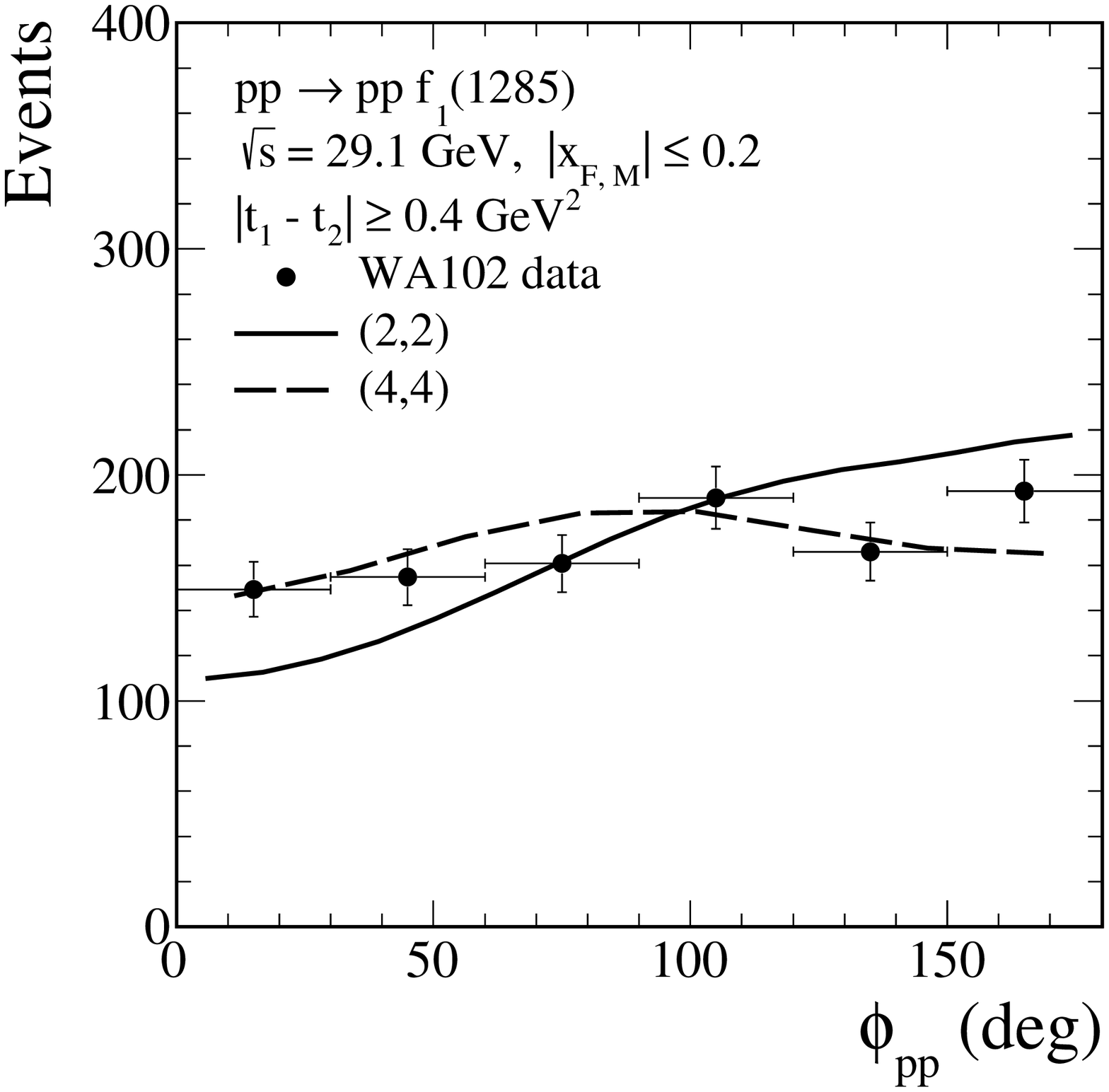}\\
\includegraphics[width=0.4\textwidth]{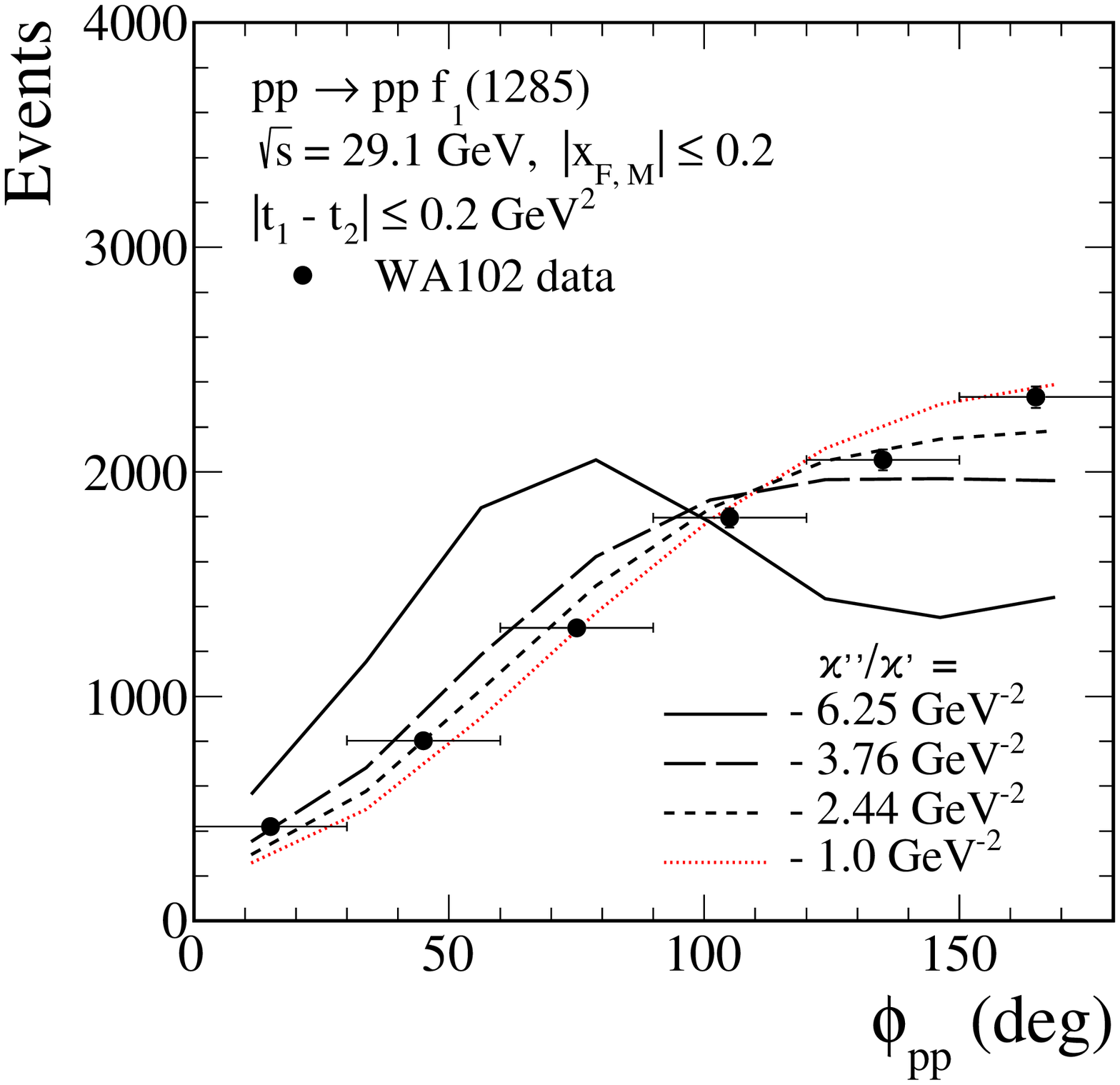}
\includegraphics[width=0.4\textwidth]{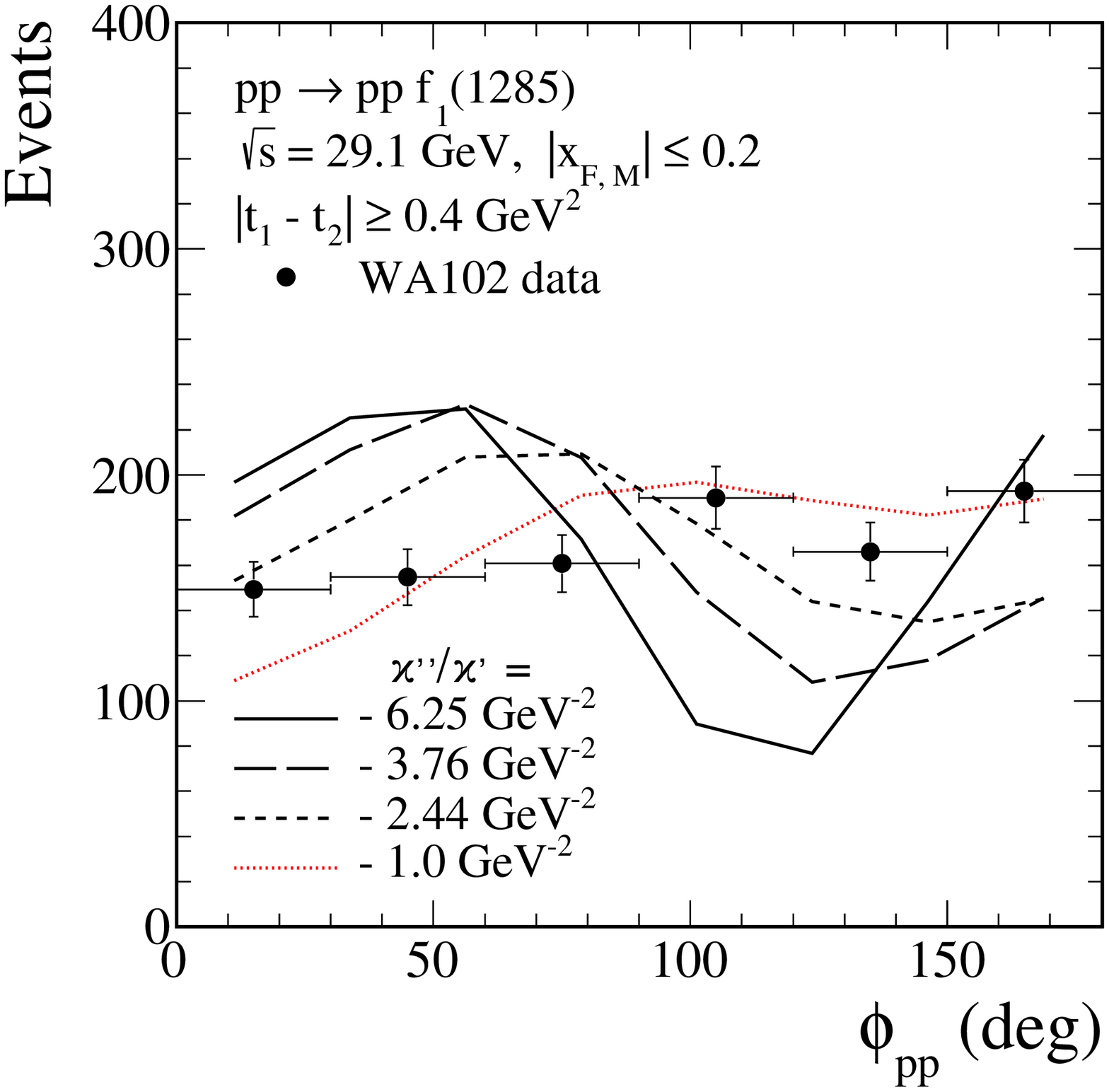}
\caption{\label{fig2}
\small
The $\phi_{pp}$ distributions for $f_{1}(1285)$ meson production
at $\sqrt{s} = 29.1$~GeV, $|x_{F,M}| \leqslant 0.2$, and
for $|t_{1} - t_{2}| \leqslant 0.2$~GeV$^{2}$ (left panels)
and $|t_{1} - t_{2}| \geqslant 0.4$~GeV$^{2}$ (right panels).
The WA102 experimental data points are from Fig.~3 of \cite{Kirk:1999df}.
The theoretical results
have been normalised to the mean value of the number of events.
The results for $\Lambda_{E} = 0.7$~GeV a form-factor parameter are shown.
}
\end{figure}

The WA102 collaboration obtained 
for the $pp \to pp f_{1}(1285)$ reaction 
the total cross section of
$\sigma_{\rm exp.} = (6919 \pm 886)\;\mathrm{nb}$
at $\sqrt{s} = 29.1$~GeV and
for a cut on the central system $|x_{F}| \leqslant 0.2$ 
\cite{Barberis:1998by}.
The WA102 collaboration also gave distributions in $t$ and
in $\phi_{pp}$ ($0 \leqslant \phi_{pp} \leqslant \pi$),
the azimuthal angle between the transverse momenta
of the two outgoing protons.
In \cite{Kirk:1999df} an interesting behaviour 
of the $\phi_{pp}$ distribution for $f_{1}(1285)$ meson production 
for two different values of $|t_{1} - t_{2}|$ was presented.
In Fig.~\ref{fig2} we show 
some of our results \cite{Lebiedowicz:2020yre}
which include - very important - absorptive corrections.
We show the $\phi_{pp}$ distribution of events
from \cite{Kirk:1999df}
for $|t_{1} - t_{2}| \leqslant 0.2$~GeV$^{2}$ (left panels) and 
$|t_{1} - t_{2}| \geqslant 0.4$~GeV$^{2}$ (right panels).
We are assuming that the reaction (\ref{1.1})
is dominated by pomeron exchange already
at $\sqrt{s} = 29.1$~GeV.
From the top panels, it seems that the $(l,S) = (4,4)$ term (\ref{2.4})
best reproduces the shape of the WA102 data.
The absorption effects play a significant role there.
In the bottom panels we examine the combination of two 
$\Pom \Pom f_{1}$ couplings
$\varkappa'$ and $\varkappa''$
calculated with the vertex (\ref{2.5}).
As discussed in Appendix~B of \cite{Lebiedowicz:2020yre},
the prediction for $\varkappa''/\varkappa'$
obtained in the Sakai-Sugimoto model is
\begin{equation}
\varkappa''/ \varkappa' = -(6.25 \cdots 2.44) \;\mathrm{GeV}^{-2}\label{kapparatiorange}
\end{equation}
for $M_\mathrm{KK}=(949 \cdots 1532)\;\mathrm{MeV}$.
This agrees with the fit 
($\varkappa''/\varkappa' = -1.0$~GeV$^{-2}$) 
as far as the sign of this ratio is
concerned, but not in its magnitude.
This could indicate that the Sakai-Sugimoto model needs a more complicated
form of reggeization of the tensor glueball propagator
as indeed discussed in \cite{Anderson:2014jia}
in the context of CEP of $\eta$ and $\eta'$ mesons.
It could also be an indication of the importance of secondary
reggeon exchanges.

We get a reasonable description of the WA102 data
with $\Lambda_{E} = 0.7$~GeV and the following possibilities:
\begin{eqnarray}
(l,S) = (2,2)\;\mathrm{term \; only}:&&
g'_{\Pom \Pom f_{1}} = 4.89\,, \;
g''_{\Pom \Pom f_{1}} = 0;
\label{3.2}\\
(l,S) = (4,4)\;\mathrm{term \; only}:&&
g'_{\Pom \Pom f_{1}} = 0\,, \;
g''_{\Pom \Pom f_{1}} = 10.31;
\label{3.3}\\
\mathrm{CS \; terms}:&&
\varkappa' = -8.88\,, \;
\varkappa''/\varkappa' = -1.0\;\mathrm{GeV}^{-2}\,.
\label{3.4}
\end{eqnarray}
Now we can use our equivalence relation (\ref{2.7})
in order to see to which $(l,S)$ couplings (\ref{3.4})
corresponds.
Replacing in (\ref{2.7}) $m^{2}$ by 
$t_{1} = t_{2} = -0.1$~GeV$^{2}$
and $k^{2}$ by $m_{f_{1}}^{2} = (1282 \;\mathrm{MeV})^{2}$
we get from (\ref{3.4})
\begin{equation}
g'_{\Pom \Pom f_{1}} = 0.42\,, \;
g''_{\Pom \Pom f_{1}} = 10.81\,.
\label{3.5}
\end{equation}
Thus, the CS couplings of (\ref{3.4}) correspond
to a nearly pure $(l,S) = (4,4)$ coupling (\ref{3.3}).

Having fixed the parameters of the model in this way 
we will give predictions for the LHC experiments.
Because of the possible influence of nonleading exchanges
at low energies, these predictions for cross sections 
at high energies should be regarded rather as an upper limit.
The secondary reggeon exchanges should give small contributions at high energies and in the midrapidity region.
As discussed in Appendix~D of \cite{Lebiedowicz:2020yre}
we expect that they should overestimate the cross sections
by not more than a factor of 4.

\subsection{Predictions for the LHC experiments}

\begin{figure}[h]
\centering
\includegraphics[width=0.42\textwidth]{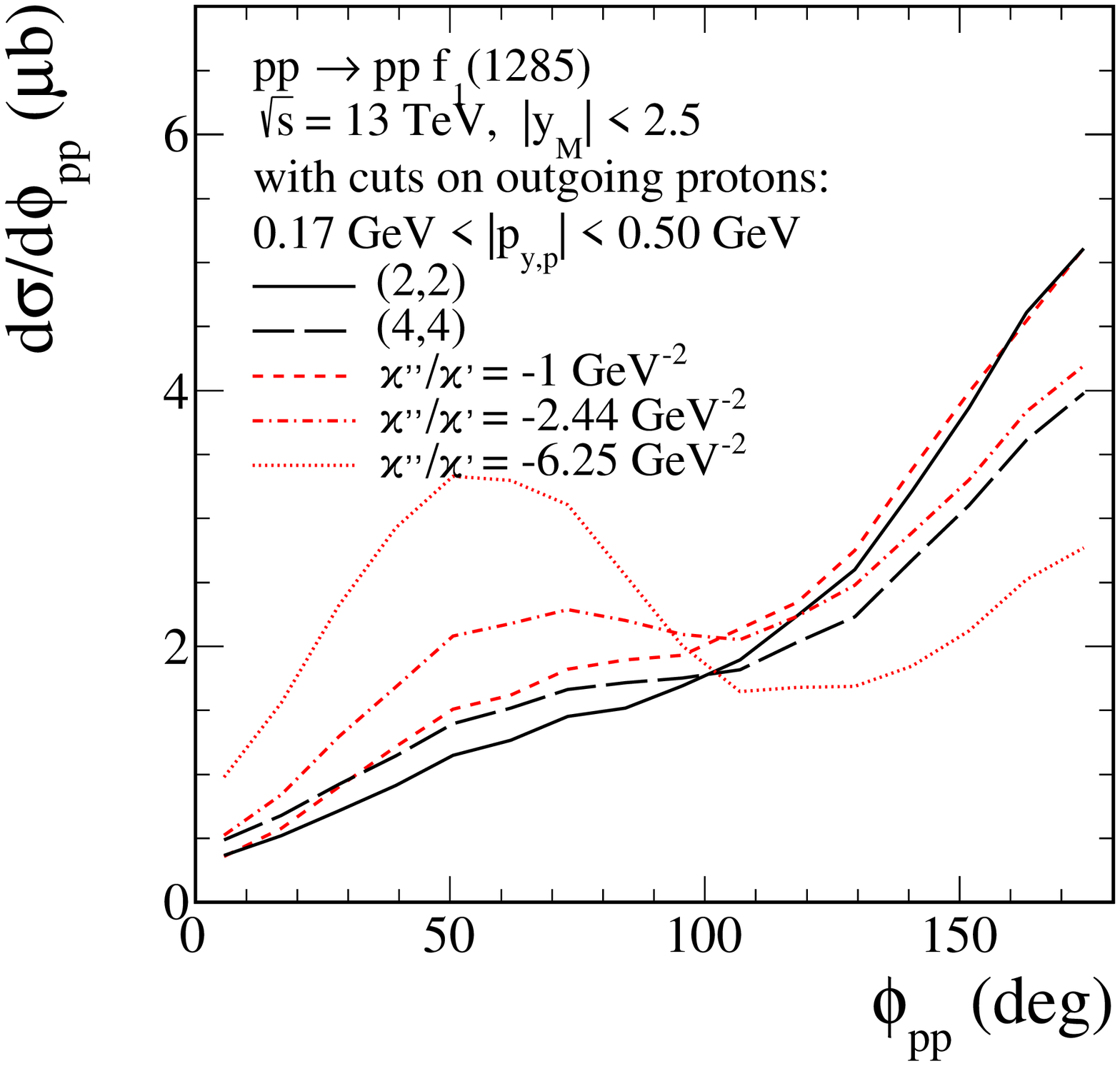}
\includegraphics[width=0.42\textwidth]{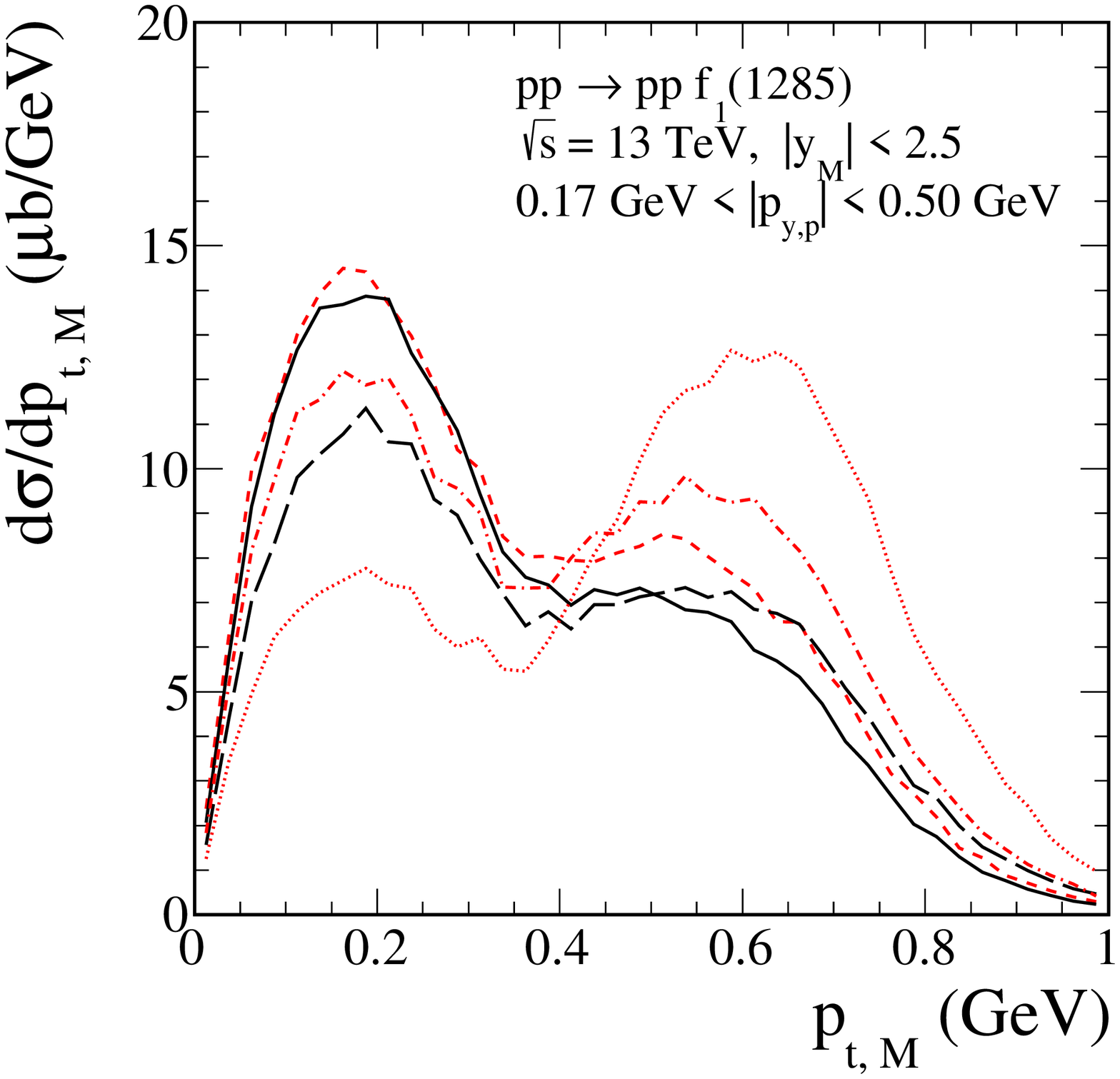}
\caption{\label{fig:ATLAS-ALFA}
The differential cross sections for
the $f_{1}(1285)$ production at $\sqrt{s} = 13$~TeV, 
$|{\rm y_{M}}| < 2.5$,
and with cuts on both outgoing protons:
$0.17\;{\rm GeV} < |p_{y,p}| < 0.50\;{\rm GeV}$.
The results for $(l,S) = (2,2)$, $(4,4)$, 
and $(\varkappa',\varkappa'')$ contributions are shown.}
\end{figure}

Now we wish to show (selected) results 
for the $pp \to pp f_{1}(1285)$ reaction for the LHC;
see \cite{Lebiedowicz:2020yre} for many more results.
In Fig.~\ref{fig:ATLAS-ALFA} we show our predictions 
for the distributions of $\phi_{pp}$
and the transverse momentum of the $f_{1}(1285)$
for $\sqrt{s} = 13$~TeV, $|{\rm y_{M}}| < 2.5$,
and for the cut on the leading protons of
$0.17\;{\rm GeV} < |p_{y,p}| < 0.50\;{\rm GeV}$.
The results for the $(l,S) = (2,2)$ term (\ref{2.3}),
the $(4,4)$ term (\ref{2.4}),
and for the $\varkappa'$ plus $\varkappa''$ terms 
calculated with (\ref{2.5}) for 
$\varkappa''/ \varkappa' = -(6.25 \cdots 2.44) \;\mathrm{GeV}^{-2}$
obtained in the Sakai-Sugimoto model 
(see Appendix~B of \cite{Lebiedowicz:2020yre}) are shown.
For comparison, the results for
$\varkappa''/\varkappa'=-1.0$~GeV$^{-2}$ are also presented.
The contribution with $\varkappa''/\varkappa' = -6.25$~GeV$^{2}$
gives a significantly different shape.
This could be tested in experiments, 
such as ATLAS-ALFA \cite{Sikora:2020mae},
when both protons are measured.
The four-pion decay channel seems well suited to measure 
the CEP of the $f_{1}(1285)$ at the LHC.
We predict a large cross section 
for the exclusive axial-vector $f_{1}(1285)$ production 
compared to the CEP of the tensor $f_{2}(1270)$ meson 
\cite{Lebiedowicz:2016ioh,Lebiedowicz:2019por}
in the $\pi^{+}\pi^{-}\pi^{+}\pi^{-}$ channel.

\section{Conclusions}
\label{sec:conclusions}

\begin{enumerate}
\item[$\bullet$] 
The calculations for the $pp \to ppf_{1}(1285)$ reaction
have been performed in the tensor-pomeron approach \cite{Ewerz:2013kda}.
We have discussed in detail the forms of the $\Pom \Pom f_{1}$
coupling. 
Detailed tests of the Sakai-Sugimoto model are possible.

\item[$\bullet$] We obtain a good description of the WA102 data at
$\sqrt{s} = 29.1$~GeV \cite{Barberis:1998by,Kirk:1999df} 
assuming that the $pp \to pp f_{1}(1285)$ reaction 
is dominated by pomeron-pomeron fusion.

\item[$\bullet$] We obtain a large cross section for CEP of
the $f_{1}(1285)$ of
$\sigma \cong 6-40 \;\mu\mathrm{b}$
for the ALICE, ATLAS, CMS, and LHCb experiments,
depending on the assumed cuts (see Table~III~of~\cite{Lebiedowicz:2020yre}).
Predictions for the STAR experiment at RHIC are given
in Table~IV of \cite{Lebiedowicz:2020yre}.
In all cases the absorption effects were included.

\item[$\bullet$] 
Experimental studies of single meson CEP reactions will allow
to extract many $\Pom \Pom M$ coupling parameters.
Their theoretical calculation is a challenging
problem of nonperturbative QCD.

\item[$\bullet$] 
Such studies could be extended, for instance by the COMPASS experiment
where presumably one could study the influence of
reggeon-pomeron and reggeon-reggeon fusion terms.
Future experiments available at the GSI-FAIR with HADES and PANDA
should provide new information about the $\rho \rho f_{1}$
and $\omega \omega f_{1}$ couplings \cite{Lebiedowicz:2021gub}.

\end{enumerate}



\paragraph{Funding information}
This work was partially supported by
the Polish National Science Centre under Grant
No. 2018/31/B/ST2/03537.





\bibliography{SciPost_Example_BiBTeX_File.bib}

\nolinenumbers

\end{document}